 \documentclass[final,3p,times]{elsarticle}
\usepackage{graphicx}
\usepackage{epsfig}
\usepackage{epstopdf}
\usepackage{times}
\usepackage{natbib}
\bibpunct{(}{)}{;}{a}{}{,}
\begin{document}

\begin{frontmatter}

\title{The extinction curve in the visible and the value of $R_V$}

\author[f1]{Frederic Zagury}

\address[f1]{Harvard University, Cambridge, MA 02138}

\begin{abstract}
This article discusses the interstellar extinction curve in the visible and  the value of the ratio of absolute to  selective extinction $R_V=A_V/E(B-V)$.
It is concluded that the visible extinction curve is likely to be linear in the visible and that indirect estimates of $R_V$ from tentative determinations of $A_V$ or from infrared and UV observations are questionable.
There is currently no evidence of any variation of $R_V$ with direction.
If $R_V$ is close to 3, as it has been inferred from mid-infrared data, starlight in the visible is extinguished by a factor $F/F_0=(2.5e^{-2\mu\mathrm{m}/\lambda})^{E(B-V)}$.
But if the visible wavelength range alone is considered, 4 appears as its most natural and probable value and $F/F_0= e^{-2E(B-V)/\lambda}$. 
\end{abstract}


\end{frontmatter}

\section{Introduction}
\begin{figure}
\resizebox{1.\columnwidth}{!}{\includegraphics{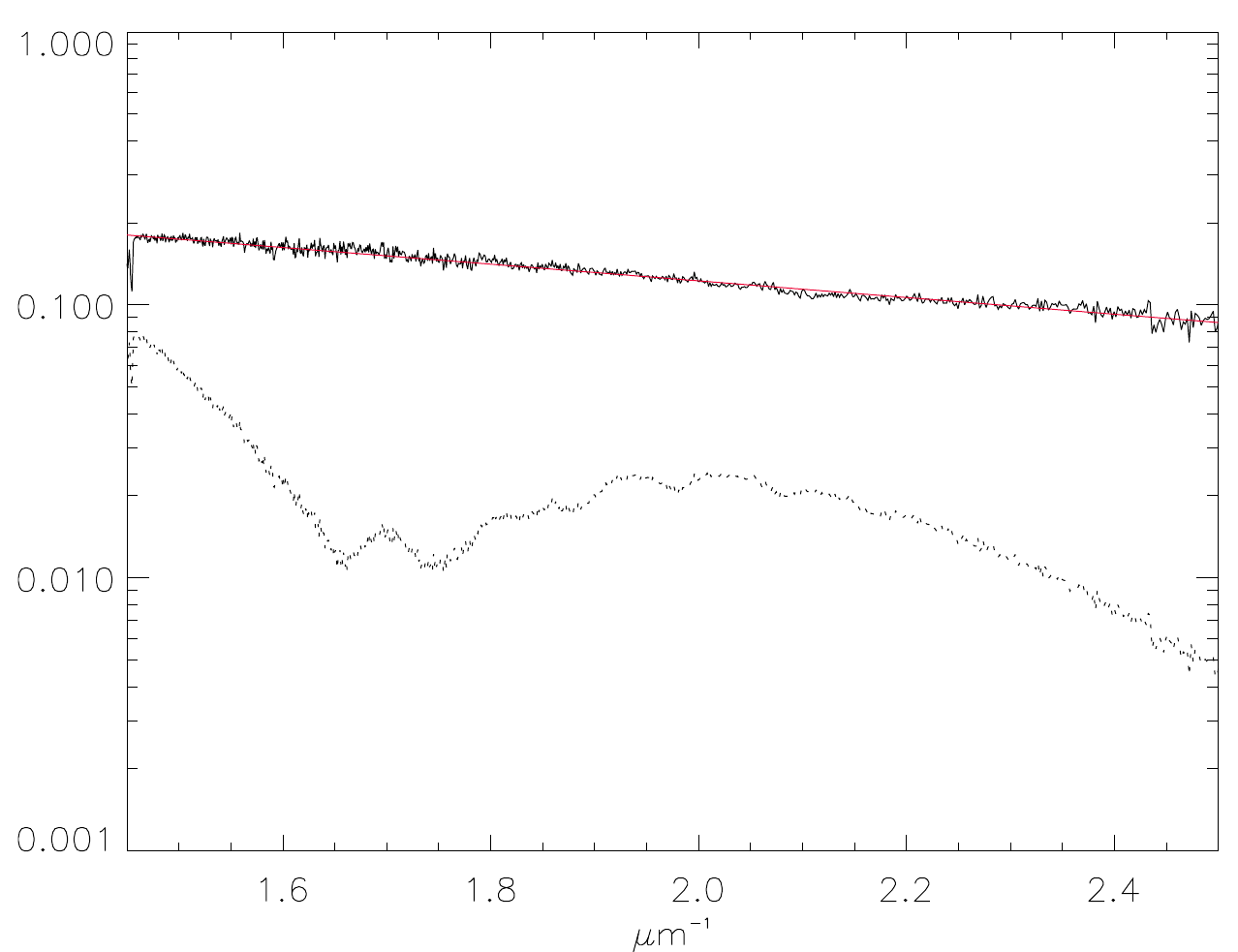}} 
\caption{Night occultation of Sirius observed by satellite GOMOS.
Bottom spectrum is the spectrum of Sirius observed at a tangent altitude of 14.8~km, normalized by the spectrum of Sirius outside the atmosphere.
Top spectrum is the same after correction for ozone and Rayleigh (nitrogen) extinctions ($N_{O_3}=3.7\,10^{20}\,\rm cm^{-2}$,  $N_{N_2}=1.2\,10^{17}\,\rm cm^{-2}$).
Only aerosol extinction remains which behaves as  an exponential of $1/\lambda$ (red trace, $\propto e^{-0.7/\lambda}$).
Courtesy of Jean-Paul Bertaux and Laurent Blanot (Service d'A\'eronomie, Verri\`eres-le-Buisson, France).
} 
\label{fig:fig1}
\end{figure}

Between 1937 and 1941 a series of papers specified the characteristics of interstellar extinction in the visible, provided a better understanding of the processes at work, and set  the general frame-work for all forthcoming studies.
 \citet{greenstein37,greenstein38} proved that extinction of a star's light in the visible was not of the Rayleigh type ($\propto e^{-a/\lambda^4}$) as previously thought\footnote{The first hint for a $\lambda^{-\alpha}$, with $\alpha$ close to 1, wavelength dependence of  interstellar extinction  came from  \citet{hall37}, in opposition to previous suggestions of a Rayleigh $1/\lambda^4$ extinction \citep{trumpler30, stebbins36}.}, but was very close to a decaying exponential of $1/\lambda$ ($\propto e^{-a/\lambda}$, that is the visible extinction law was linear in $1/\lambda$).
Such laws are common in nature (atmospheric aerosols, Fig.~\ref{fig:fig1}, are the best but not the only example), and attributed to size distributions of particles, with no specific implication on their chemical composition.
These distributions are usually associated with large albedos and  strong forward scattering phase functions, two properties which in the case of interstellar dust were confirmed in 1940  \citep{henyey40,henyey41}.

Linear extinction laws are defined by two parameters, their slope, proportional to the reddening $E(B-V)=A_B-A_V$ on the line of sight, and the extinction at a specific wavelength $A_{\lambda_0}$.
The extinction at wavelength $\lambda$ normalized by the reddening, $A_\lambda/E(B-V)$,  depends on the absolute extinction to color excess ratio ($R_V$ in the Johnson photometric system) alone (Sects.~\ref{def} and \ref{evis}).

Greenstein found good spatial homogeneity of interstellar extinction.
Assuming that the extinction law goes to 0 in the infrared  \citep{stebbins39} a value of $R_V=4$ is obtained (see also \citet{mendoza65}).
In contrast \citet{greenstein41} found from a statistical analysis of the mean reddening at two different wavelengths that interstellar extinction becomes zero at 3.2~$\mu$m, which implies $R_V=3.3\, (\pm 0.1)$.

Since the 1940s  the debate on interstellar extinction has focused on four essential questions.
How close to a perfect exponential is the extinction of starlight in the visible? 
Does extinction vary from direction to direction (from an interstellar cloud to another)? 
On how many parameters does it depend? 
How does it constrain the composition of interstellar dust?
 
The purpose of this paper is to evaluate which logical conclusions on interstellar extinction in the visible and the near-infrared  ($4300\,\rm\AA - 1.4\,\mu$m) may be reached from an analysis of the abundant and often contradictory literature on the subject:
what is the shape of the visible extinction curve?
Is the key parameter $R_V$ a constant or is it highly variable, as different studies separately claim, and what is its value(s)?
Can extinction in the UV or in the infrared be related to the visible extinction and to the value of $R_V$?

Sect.~\ref{def}  gives the relevant definitions that are used in this paper.
Sects.~\ref{evis} to \ref{euv} review the main features of interstellar extinction in  three wavelength domains which have been used to investigate the value of $R_V$, the visible, the infrared (near and mid-infrared, $0.8\mu\rm m\leq\lambda\leq 3\mu$m), and the UV.
Sect.~\ref{rv}  questions the variability  of $R_V$.
Sect.~\ref{dis} attempts to give the most logical responses to the questions raised in the previous paragraph.
\section{Definitions} \label{def}
Let $F_\lambda$ be the flux of a reddened star, measured on earth, and $F_{0,\lambda}$ that of an unreddened star of same spectral type.
The quantity which is readily accessible through observation is the excess of extinction of the reddened star at wavelength $\lambda$ over the extinction at a given wavelength $\lambda_0$,
\begin{equation}
A_\lambda-A_{\lambda_0}=E(\lambda-\lambda_0)=-2.5\log\frac{\frac{F_\lambda}{F_{0,\lambda}}}{\frac{F_{\lambda_0}}{F_{0,\lambda_0}}} 
 \label{eq:aext}
\end{equation}
Extinction in different directions can be normalized to the same amount of reddening between $\lambda_0$ and another given wavelength $\lambda_1(<\lambda_0)$ by dividing  $E(\lambda-\lambda_0)$ by $E(\lambda_1-\lambda_0)$.
Normalized extinction curves are defined by
\begin{equation}
g_{\lambda_1,\lambda_0}(1/\lambda)=\frac{E(\lambda-\lambda_0)}{E(\lambda_1-\lambda_0)}
=\frac{A_{\lambda}}{E(\lambda_1-\lambda_0)}-R_{\lambda_1,\lambda_0}
 \label{eq:ec}
\end{equation}
with
\begin{equation}
R_{\lambda_1,\lambda_0}=\frac{A_{\lambda_0}}{E(\lambda_1-\lambda_0)}
 \label{eq:r}
\end{equation}
$g_{\lambda_1,\lambda_0}(1/\lambda)$ measures the extinction per unit amount of interstellar matter along the line of sight.

Normalized extinction curves  $g_{\lambda_1,\lambda_0}$ in different directions all pass through the points $(1/\lambda_0, 0)$ and $(1/\lambda_1,1)$ in the $(1/\lambda,g_{\lambda_1,\lambda_0})$ plane, and  have the same mean slope,  $(\lambda_0^{-1}-\lambda_1^{-1})^{-1}$.
For extinction to be the same in different directions normalized extinction curves need to be identical.
In this case knowledge of $E(\lambda_1-\lambda_0)$  fixes the amount of extinction and the quantity of interstellar dust on the line of sight at all wavelengths.
If $g_{\lambda_1,\lambda_0}$ varies with direction extinction depends on E(B-V) and one or more additional parameters. 

Two systems of normalization, $(\lambda_3,\lambda_2)$ and $(\lambda_1,\lambda_0)$ can be converted one into the other by
\begin{equation}
g_{\lambda_3,\lambda_2}(1/\lambda)= \frac{g_{\lambda_1,\lambda_0}(1/\lambda)-g_{\lambda_1,\lambda_0}(1/\lambda_2)}
{g_{\lambda_1,\lambda_0}(1/\lambda_3)-g_{\lambda_1,\lambda_0}(1/\lambda_2)}       \label{eq:conv}
\end{equation}
In Nandy's system \citep{nandy64}, $1/\lambda_0=2.22~\rm\mu m^{-1}$ and $1/\lambda_1=1.22~\rm\mu m^{-1}$ were chosen to give a mean slope of unity.
In the  \citet{johnson53} photometric system ($1/\lambda_0=1/\lambda_V=1.82~\rm\mu m^{-1}$, $1/\lambda_1=1/\lambda_B=2.27~\rm\mu m^{-1}$)  the slope  is 2.22, and normalized extinction curves are defined by
\begin{equation}
g_0(1/\lambda)=\frac{A_{\lambda}}{E(B-V)}-R_V
 \label{eq:g0}
\end{equation}
with
\begin{equation}
R_V=\frac{A_V}{E\left(B-V\right)}=\frac{1}{\frac{\sigma_B}{\sigma_V}-1} \label{eq:rv}
\end{equation}
$\sigma_B$ and $\sigma_V$ are the extinction cross-sections of interstellar dust in the B and V bands in the direction of the observation.

Eq.~\ref{eq:g0} emphasizes the importance of $R_V$.
If $R_V$ is known, $A_\lambda$ becomes measurable at all wavelengths.
Eq.~\ref{eq:rv}  shows that $R_V$ depends only on the extinction properties  of  interstellar dust  in the B and V bands (along the line of sight).
Identical normalized extinction curves in different directions certainly imply equality of $R_V$ in these directions.

The constancy or not of $R_V$ and its exact value(s) determine our ability to evaluate $A_\lambda$ in any direction, to correct observations of far-away objects for foreground reddening, and to parametrize interstellar dust models. 
The  \citet{schlegel98} derivation of $A_\lambda$ in different photometric bands from $E(B-V)$ for instance relies on the assumption that $R_V$ is constant and equal to 3.1; their estimate of interstellar extinction in any direction is widely used today.

There is no direct and straightforward way to estimate $R_V$.
$E(B-V)$ is deduced from a star spectral type and flux ratio between the B and V bands but $A_V$ can not be ascertained unless the star absolute luminosity and  distance are precisely known.
Existing determinations of $R_V$  rely on indirect methods,  the extension of  $g_0$ to the infrared (Sect.~\ref{enir}), estimates of $A_V$, or on the UV extinction curve (Sect.~\ref{rv}).
\section{The visible extinction curve} \label{evis}
\begin{figure}
\resizebox{1.\columnwidth}{!}{\includegraphics{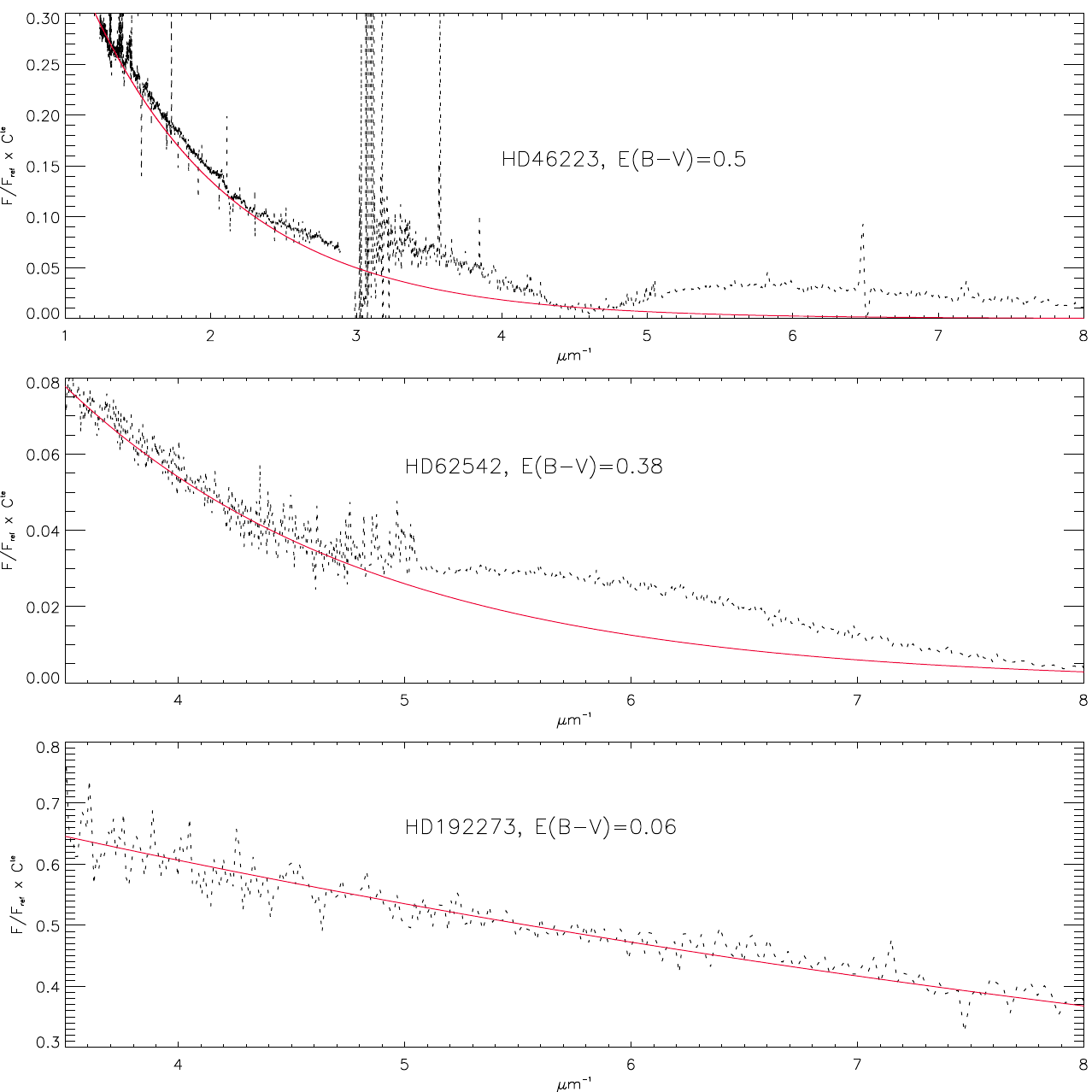}} 
\caption{Extinction of HD46223, HD62542, HD192273.
From top to bottom,  ratios of the spectra of HD46223 to the unreddened star of same type HD269698 (15~Mon), of HD62542 to HD32630,  of HD192273 to HD31726.
The red lines are the exponentials $e^{-2 \Delta (B-V)/\lambda}$, with $\Delta(B-V)$ the B-V difference between the reddened and the reference stars.
The extinction of  HD192273 is linear throughout the visible-UV wavelength range, in the visible to the near-UV for HD62542, and in the visible only for HD46223.
The $F/F_0$ ratios have been normalized so that the  fits be exactly $e^{-2 \Delta (B-V)/\lambda}$ (which implies $R_V=4$, Sect.~\ref{dis}).
} 
\label{fig:fig2}
\end{figure}
With the exception of  Divan's work \citep{divan54} spectral studies of interstellar extinction in the visible are rare.
There is a lack of a large, reliable data-base of stellar spectra free from atmospheric extinction, such as  exists in the UV (the IUE data-base).
Spectra in analytical form did not become common before the 1950-1960s and a precise
removal of atmospheric extinction (which also has a dependence  on $1/\lambda$, Fig.~\ref{fig:fig1}) is not obvious.
Photometry, sometimes from only a few data-points,  has been the standard means to study extinction in the visible, although it is less accurate than spectral analysis can be.

The upper plot of Fig.~\ref{fig:fig2} is the ratio $F/F_0$, in the [1250~\AA, 8000~\AA] wavelength range, for the star HD46223\footnote{observations from J.F. Le Borgne library of stellar spectra, http://www.ast.obs-mip.fr/article181.html \citet{leborgne03}}.
With an excellent precision the visible part of the spectrum follows a decaying exponential of $1/\lambda$ (and diverges from it in the UV).

This exponential decrease of  the extinction of starlight in the visible is confirmed by most studies\footnote{two notable exceptions are the photometric observations of Johnson (fig.~12 in \citet{johnson65}) and of  \citet{schild77} who finds that '\emph{in no octave of the spectrum is the interstellar absorption strictly proportional to reciprocal wavelength.....it is becoming increasingly clear that  variations with Galactic longitude are important}' (see also \citet{borgman54,johnson55,johnson68}). Johnson's observations and methods (especially his summary in \citet{johnson68})  are  criticized in several articles \citep{ sherwood75,rozis56, schalen75, schultz75}.} of visible normalized extinction curves.
 \citet{divan54},  \citet{whitford58},  \citet{nandy64},  \citet{ardeberg82} and more recently \citet{bondar06} found the same  $g_{\lambda_1,\,\lambda_0}(1/\lambda)$  in all directions (over the visible domain).  

\begin{figure}
\resizebox{\columnwidth }{!}{\includegraphics{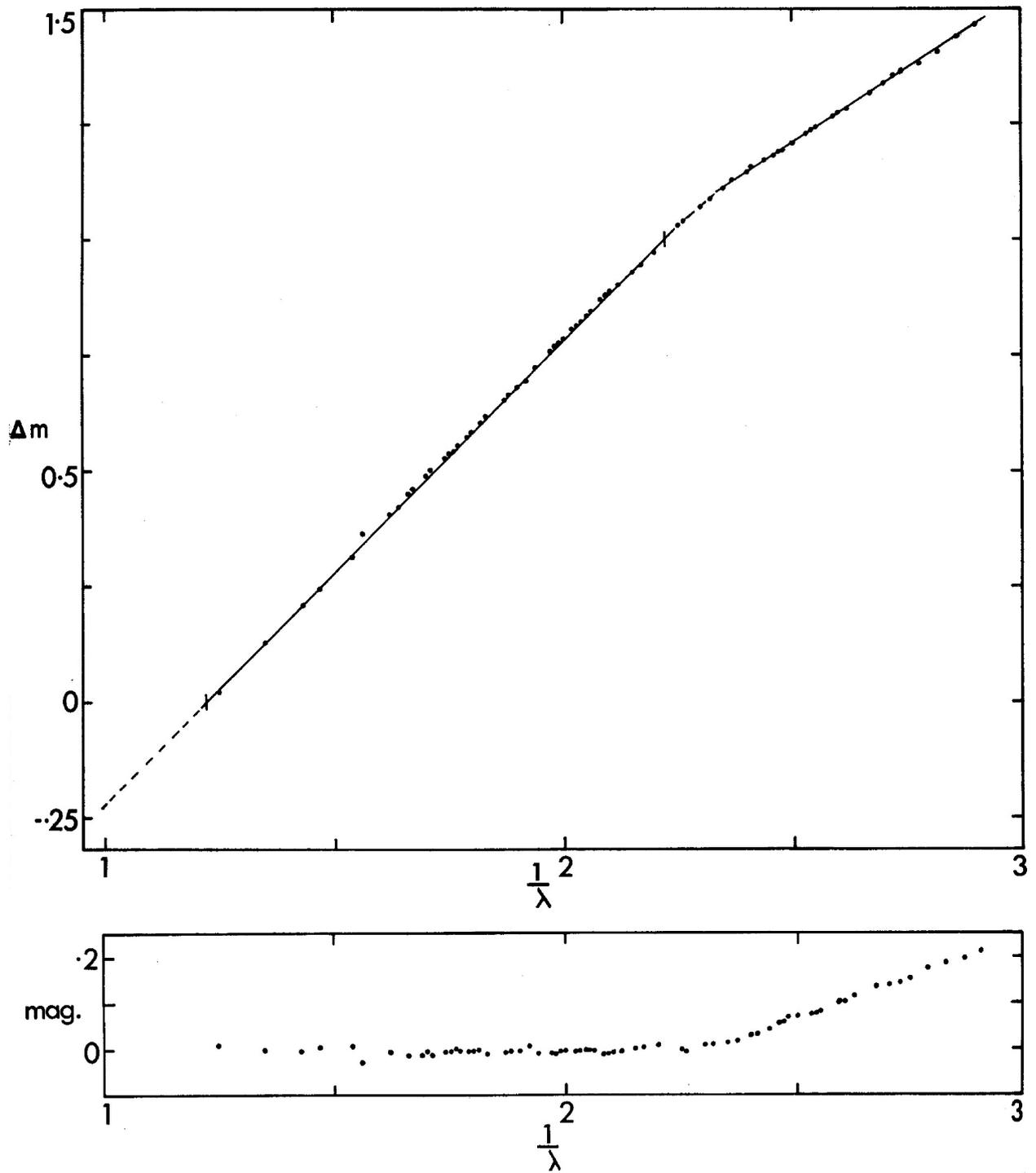}} 
\caption{The visible normalized extinction curve (top) derived by K.~Nandy, and the error margin  (figs~14 and 15 in \citet{nandy64}). 
The curve is linear down to $1/\lambda\sim 0.8\,\rm \mu m^{-1}$ ($\lambda\sim1.2\,\mu$m, see figs.~1 and 2 in \citet{whitford58}, fig.~5 in \cite{hoyle69}, and  fig.~2 in \citet{ardeberg82}).
} 
\label{fig:fig3}
\end{figure}

The most extensive data-set of extinction curves in the visible was gathered by Nandy in the 1960-1970s (see review, \citet{wick98}).
The extinction curve weighted over a large number of observations that he derived in 1964 \citep{nandy64} is reproduced in Fig.~\ref{fig:fig3}.
It is  to a high degree of accuracy linear in the visible from 2.3~$\mu\rm m^{-1}$ (4350~\AA) to 0.8~$\mu\rm m^{-1}$ ($1.2\,\mu$m) in the near-infrared \citep{wick98, whitford58}.
In this wavelength range the exponential dependence of $F_\lambda/F_{0,\lambda}$ can equally be expressed as
\begin{equation}
\frac{F_\lambda}{F_{0,\lambda}}= C_0e^{-a/\lambda}= e^{-(a/\lambda+b)} 
 \label{eq:f}
\end{equation}
or by
\begin{eqnarray}
 A_\lambda &\,=\,&
 \frac{ \left(A_B-A_V\right)}{\frac{1}{\lambda_B}-\frac{1}{\lambda_V}}
 \left(\frac{1}{\lambda}-\frac{1}{\lambda_V}\right) 
 +A_V \nonumber \\
&\,=\,& 2.22E\left(B-V\right)\left(\frac{1\,\mu{\rm m}}{\lambda}+0.46\left(R_V-4\right)\right)  \label{eq:al}
\end{eqnarray}
From Eqs.~\ref{eq:f} and \ref{eq:al}
\begin{eqnarray}
a&= &2.02E(B-V) \label{eq:a}
 \\
b&=&0.92E(B-V)\times (R_V-4)  \label{eq:b}
\end{eqnarray}
The extinction per unit reddening is
\begin{eqnarray}
\frac{ A_\lambda}{E\left(B-V\right)} &\,=\,& 2.2\left(\frac{1\,\mu{\rm m}}{\lambda}+0.46\left(R_V-4\right)\right)  \label{eq:aln}
\\
g_0(\lambda)&\,=\,& \frac{2.2\,\mu{\rm m}}{\lambda}-4 \label{eq:g0l} \\
\left(\frac{F_\lambda}{F_{0,\lambda}}\right)_{E\left(B-V\right)=1}&=&e^{-0.92(R_V-4)}e^{-2\mu{\rm m}/\lambda} \label{eq:fn}
\end{eqnarray}
\section{The infrared extinction curve} \label{enir}
Observed extinction curves are linear only in the visible and part of the near-infrared domain.
Fig.~\ref{fig:fig3} shows that the curve deviates in the near-UV above $1/\lambda\sim 2.3\,\rm\mu m^{-1}$ (shortly after the $B$-band) where there is less extinction than expected.
The error margin on Fig.~\ref{fig:fig3} increases dramatically because extinction in the UV no longer depends on $E(B-V)$ alone (Sect.~\ref{euv}). 

In the infrared extinction is  much less than at shorter wavelengths.
Error margins are large and the extinction curve is difficult to ascertain.
Infrared extinction data are limited to a few photometric bands, and thermal emission from dust grains shows up above  a few $\mu$m (\citet{sherwood75} and references therein).
Below $0.8\,\rm \mu m^{-1}$ published mid-infrared extinction laws  do not follow the visible law (\citet{stead09}, figs.~1 and 2 in \citet{whitford58}, fig.~2 in \citet{ardeberg82}).
The extinction curve seems to flatten shortward of $1/\lambda\sim0.8\,\rm\mu m^{-1}$ ($1.2\,\rm\mu m$) and extinction may be larger  than predicted by the linear behavior in the visible.

Mid-infrared extinction was used   \citep{johnson65, johnson68, wegner03} to probe variations of $R_V$   ($g_0(1/\lambda\rightarrow 0)\rightarrow -R_V$, Eq.~\ref{eq:g0}).
These findings are contradicted by several studies \citep{bondar06, sherwood75, rozis56, schalen75, schultz75}, sometimes even in the same regions of the sky.
There is today no tangible proof of spatial variations of the extinction in the infrared. 
The legitimacy of using the infrared to determine $R_V$ is also questionable, especially  with respect to the spatial constancy of $g_0(\lambda)$ in the visible wavelength range (Sect.~\ref{dis} and \citet{mendoza65}).
\section{The UV extinction curve} \label{euv}
The  UV extinction curve exhibits the well-known  2200~\AA\ bump feature (upper plot of Fig.~\ref{fig:fig2}); the extinction is less  than expected from the continuation of the visible extinction curve (Figs.~\ref{fig:fig2}). 
Early in the 1970s it was recognized that normalized $g_0(1/\lambda)$ curves depend on direction: $E(B-V)$ alone is not enough to determine the whole extinction curve. 
These properties  have largely contributed to the development of new interstellar dust models (see \citet{hoyle91}). 

\citet{ccm89} (CCM in the following)  proved in 1989 that most normalized curves with a bump can be deduced from a function of $1/\lambda$ which depends upon a single parameter, $R_{ccm}$. 
In the CCM framework the size of the bump depends on $E(B-V)$ only (this is a crude approximation, see \citet{savage75}) while the free parameter determines the average slope of the extinction curve (with respect to the one in the visible) and its shape in the far-UV.
\citet{ccm89}   assume that $R_{ccm}$ is equal to $R_V$, an affirmation to be discussed in Sect.~\ref{rv}.

Cases also exist where the linear visible extinction extends to the near-UV, eventually to the far-UV (Fig.~\ref{fig:fig2}).
A linear extinction curve over all the visible-UV can be found in the directions of very low column density within the Galaxy  (Fig.~\ref{fig:fig2}, bottom plot, and \citet{z01}); similar extinction curves are observed in the Magellanic Clouds along lines of sight with even higher column density \citep{mc}.
In some directions, HD62542 (Fig.~\ref{fig:fig2}, middle plot) or HD29647 in our Galaxy, Sk-69228 and Sk-70116 in the LMC \citep{mc}, the  extinction is linear down to the bump region and diverges from linearity in the far-UV only.
All these directions have no bump and  cannot be fitted by the CCM function.
The CCM fit can be improved to include linear extinction curves \citep{mc}.
\section{On the variations of $R_V$ } \label{rv}
\begin{figure}
\resizebox{1.\columnwidth}{!}{\includegraphics{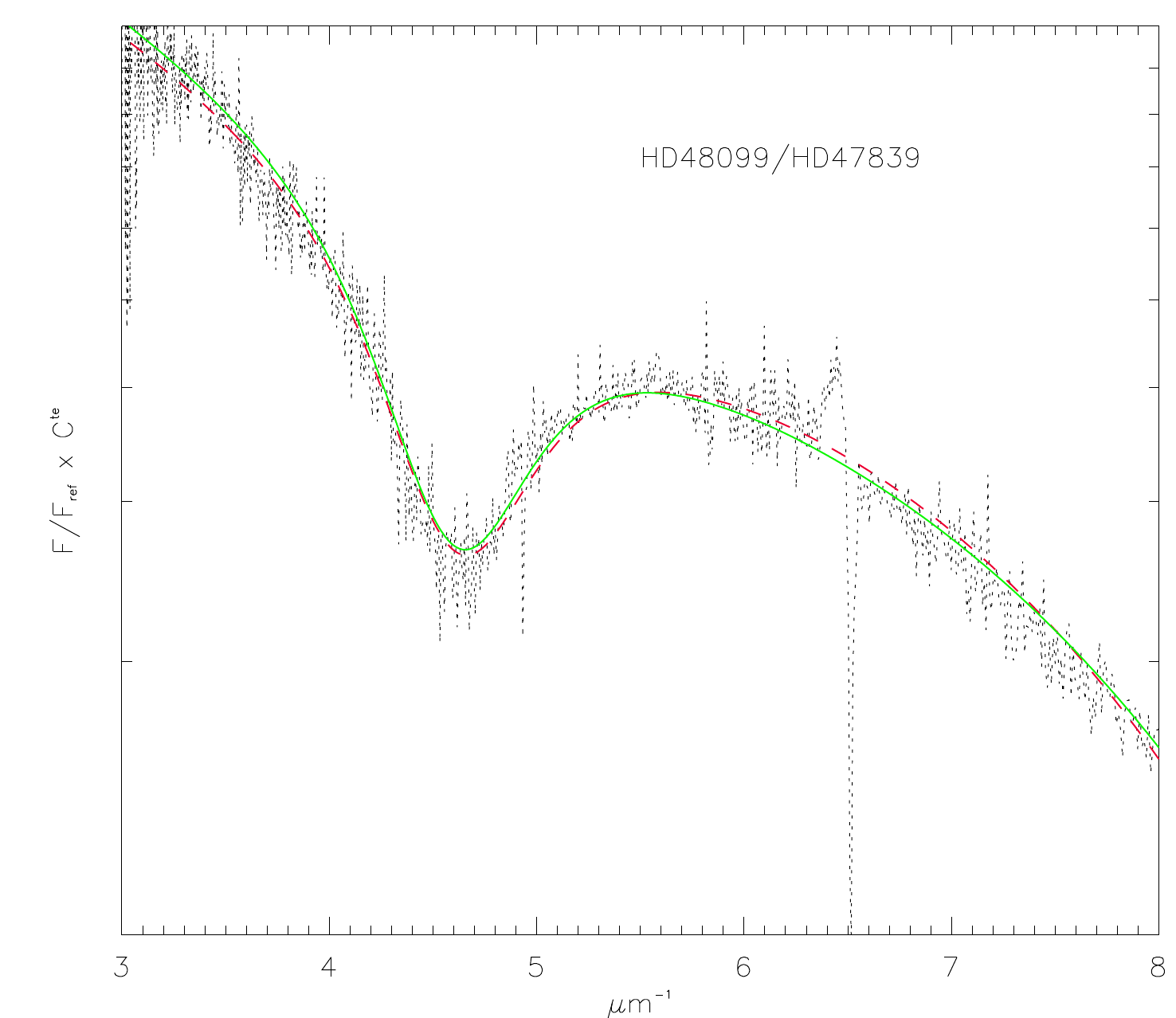}} 
\caption{Dotted spectrum is the ratio of HD48099 (reddened star) to the slightly reddened  ($E(B-V)=0.07$) star HD47839, corrected for extinction.
If the slight reddening  ($E(B-V)=0.07$) of HD47839 is considered, the CCM fit is the red dashed spectrum, with free parameter $R_{ccm}=1.2$. 
If half only of this reddening is corrected, the CCM fit is the green spectrum, with  $R_{ccm}=3.4$, close to the value found in \citet{ccm89}.
A slight error in the adopted value of $E(B-V)$ can thus lead to largely different values of the CCM free parameter.
The plot also shows that $R_{ccm}$ has no impact on the size of the bump, which, in the CCM framework, is controlled by the reddening $E(B-V)$.
} 
\label{fig:fig4}
\end{figure}
If, in the visible, normalized extinction curves  $g_0(1/\lambda)$ do not depend on direction (Sect.~\ref{evis}) there is no reason why  the relative extinction cross-sections of interstellar grains in the R and B bands (Eq.~\ref{eq:rv}) should vary, and $R_V$ should remain constant (Sect.~\ref{def}). 

Paradoxically the constancy of $R_V$ is disputed not so much from the study of interstellar extinction in the visible itself  but from indirect and generally more speculative methods, which  often involve the extinction in other wavelength ranges. 
They rely either on infrared data (which are not conclusive, Sect.~\ref{enir}); on tentative estimates of $A_V$; or on the CCM fit of UV extinction curves (Sect.~\ref{euv}). 

Direct determination of $A_V$  uses star count methods or open clusters  distance estimates.
Star counts for the same region have resulted to contradictory conclusions regarding the nature and variability of $R_V$ \citep{johnson68, schalen75, balazs04}.
The values of $R_V$ range from 3 to 6 \citep{balazs04}. 
Distance estimates of open clusters are no more convincing.
In a recent article \citet{turner11} shows that the distance to open cluster Shorlin~1 should be revised from over 10~kpc to less than 3~kpc and thereby demonstrates the large errors which are associated to distance estimates by means of photometry.
The same paper also  lists some thirty determinations of $R_V$  in the Carina region which  range from 3 to 5.2.
Such variations over  a small area of the sky are obviously difficult to reconcile with the constancy of $g_0$ found by Nandy, Divan, Bondar, and others (Sect.~\ref{evis}).

The same remark applies to the UV.
The  use (since 1989) of  the CCM fit  to probe large and local variations of $R_V$ is questionable. 
The  free parameter $R_{ccm}$ of an extinction curve is obtained from eq.~1 in CCM
\begin{equation}
\frac{A_\lambda}{A_V}= a(x) + \frac{b(x)}{R_{ccm}},
 \label{eq:ccm}
\end{equation}
where $a(x)$ and $b(x)$ are functions of $x=\lambda^{-1}$ alone \citep{ccm89}.
For the B band ($x_B=2.27\rm\mu m^{-1}$) Eq.~\ref{eq:ccm} yields, in any direction
$$
\frac{A_B}{A_V}= a(x_B) + \frac{b(x_B)}{R_{ccm}},  \nonumber\\
$$
or
$$
\frac{1}{R_V}= a(x_B)-1 + \frac{b(x_B)}{R_{ccm}}  \nonumber
$$
If $R_{ccm}=R_V$
\begin{equation}
R_V=-\frac{1-b(x_B)}{1-a(x_B)}.
  \label{eq:rvccm}
\end{equation}
Should $R_{ccm}$ be equal to $R_V$  Eq.~\ref{eq:rvccm} shows that it would have to be a constant (from Eq.~\ref{eq:rvccm} and eqs.~3a and 3b in CCM: $R_V= 1.48$) independent of direction.
This is in contradiction with the fact that the free parameter of the CCM parameterization, $R_{ccm}$, varies according to direction.
$R_V$ is therefore unlikely to be the free parameter of the CCM function.

It is further difficult to conceive how $R_{ccm}$, which in practice fixes the average UV slope of the CCM fit  with respect to the mean slope of  $g_0(1/\lambda)$ in the visible  (fig.~4 in \citet{ccm89} and  \citet{mc}), can influence the visible extinction curve, let alone $R_V$.
It  also should be noted that the parameter $R_{ccm}$ is highly sensitive to the estimated value of $E(B-V)$  (Fig.~\ref{fig:fig4}), and is not necessarily the same whether the CCM law or its improved formulation in \citet{mc} are employed.
Last, $R_{ccm}$  barely agrees with the value of $R_V$ derived by other methods, which find, using similar data-sets, either different values in the same region (in the Magellanic Clouds for instance \citep{mc}) or  a constant $R_V$ \citep{bondar06,schlafly11}.
\section{Discusssion} \label{dis}
The interstellar extinction curve is now accessible through the entire spectrum although studies in each of
the three wavelength regions (the infrared, the visible, and the UV) are generally done separately and  rely on observations with different quality, sensitivity, and resolution. 
In the infrared there is no conclusive evidence that interstellar extinction should depend on another parameter in addition to $E(B-V)$ (Sect.~\ref{enir} and \citet{sherwood75}).

Extinction in the UV, although the best documented, remains to be understood.
The \citet{ccm89} paper had a strong impact and was a remarkable breakthrough.
It proved that interstellar extinction in the UV depends on $E(B-V)$ and on an additional parameter only.
The CCM function, representative of  normalized extinction curves with a bump at 2200~\AA, is however purely empirical and void of  physical significance.
Its free parameter is unlikely to be $R_V$ (Sect.~\ref{rv}).
The nature of the parameter which governs UV extinction is still to be determined. 

Over sixty years of observations support a linear extinction law in the visible, with $g_0(1/\lambda)$ normalized extinction curves independent of direction  (Sect.~\ref{evis}).
Linear extinction laws are attributed to grain size distributions ($\Phi(a)da= a^{-4}da$ in \citet{vdh}).
They have no  implication for the chemical composition of the grains responsible for the extinction: extinction by aerosols in the atmosphere, like interstellar dust, follows a similar law.
The extinction at a specific wavelength is due to particles of size close to the wavelength.

The spatial uniformity of $g_0(1/\lambda)$ in the visible strongly favors a constant value of $R_V$ (Sect.~\ref{def}).
Estimates of $A_V$ in different directions, which essentially are of a statistical character (\citet{balazs04} and Sect.~\ref{enir}), suggest that $R_V$ should be between 3 and 6.

The flattening of the extinction curve observed in the near-infrared (fig.~2 in \citet{whitford58}), which interrupts the linear decrease of $g_0(1/\lambda)$ towards the longest wavelengths, was used to derive a value of $R_V$ close to 3.
If $R_V$ is indeed close to 3, Eq.~\ref{eq:fn} may be rewritten as
\begin{equation}
\left(\frac{F_\lambda}{F_{0,\lambda}}\right)_{E\left(B-V\right)=1}=2.5e^{-2\mu{\rm m}/\lambda}
 \label{eq:rv3}
\end{equation}
It is however arbitrary to determine $R_V$ from infrared data since below 0.8~$\mu\rm m^{-1}$  the interstellar extinction curve seems not to follow the same analytical law as in the visible.
$R_V$ should depend  on the distribution of those interstellar dust grains with sizes close to $\lambda_B$ and $\lambda_V$ and be determined from the extension to longer wavelengths of  $A_\lambda/E(B-V)$ rather than from the observed mid-infrared data.
The mid-infrared flattening, should it be confirmed, may as well result from a process  independent from the extinction at shorter wavelengths, from a change in grain-size distribution for the largest interstellar grains for instance.

A general difficulty in discussing these estimates of $R_V$ is that  $R_V$ can be defined in different ways which do not necessarily have the same meaning.
The exact definition, $R_V=A_V/E(B-V)$, involves only the reddening in the $B$ and $V$ bands.
Because direct determination of these quantities is particularly difficult  alternative definitions rely on the limit of $g_0(1/\lambda\rightarrow 0)$, and on the supposition that $A_\lambda(1/\lambda\rightarrow 0)\rightarrow 0$ which is true only if there is no gray (neutral) extinction.
But the $g_0(1/\lambda\rightarrow 0)$ limit can also be considered in two ways, either  by taking the limit of mid-infrared observations, as it is generally done, or by continuing Nandy's linear extinction law (Fig.~\ref{fig:fig3}) to the longer wavelengths.

As stated above the infrared method, which  leads to $(R_V)_{IR}\approx 3$ and to Eq.~\ref{eq:rv3}, is not consistent with the original definition of $R_V$.
If there is a flattening of the extinction in the near-infrared $(R_V)_{IR}$ must be smaller than $A_V/E(B-V)$ and underestimates $R_V$.

If Nandy's curve in the visible alone is considered, provided that  the $B$-band lies within the limits of the linear part of normalized extinction curves (that is, if $A_B$ is not or little affected by the near-UV departure from linearity, Fig.~\ref{fig:fig3}), and assuming no gray extinction, Eqs.~\ref{eq:g0} and \ref{eq:g0l} necessarily imply $R_V=4$.
This conclusion can be re-formulated in the following way.
If linear normalized extinction curves in the visible  were to be considered alone one would expect, from any size distribution of particles, that the coefficient in front of the exponential in Eq.~\ref{eq:rv3}  be less than 1 (instead of 2.5).
In absence of neutral extinction, which is presumed to be the case for interstellar dust  \citep{dufay54,sherwood75}, this coefficient, $C_0$ in Eq.~\ref{eq:f}  (or $e^{-0.92(R_V-4)}$ in Eq.~\ref{eq:fn}), should be 1.
Then
\begin{eqnarray}
\frac{F_\lambda}{F_{0,\lambda}}&=&e^{-\frac{2\mu{\rm m}}{\lambda}E\left(B-V\right)} \label{eq:fn4}\\
\frac{ A_\lambda}{E\left(B-V\right)} &\,=\,& \frac{2.2\,\mu{\rm m}}{\lambda}  \label{eq:alf}
\\
R_V&=&4 \label{rv4}
\end{eqnarray}
$R_V=4$ is precisely the value \citet{turner11} has  obtained from his most recent photometric measurements in the Carina region.
This value is not "anomalous" as it was often suggested, but rather appears as a logical outcome of the linearity of interstellar extinction in the visible. 

This paper was motivated by the large and logically incompatible discrepancies that appear from a survey of the existing literature on the determination of $R_V$.
Depending on the study $R_V$ has been found to be either constant and independent of the line of sight or highly variable from one direction to another.

It is my conclusion that the normalized visible extinction law is most likely linear and independent of direction so that $R_V$ should be the same constant for all directions.
Although a value close to 3 is generally preferred, $R_V=4$ would more naturally fit with the linear extinction law found in the visible.
\section*{Acknowledgments}
This work was funded by an Arthur Sachs  fellowship. I am grateful for the hospitality and resources provided by Harvard  University.

{}
\newpage

\end{document}